\documentclass[prb,twocolumn,superscriptaddress]{revtex4-1}
\pdfoutput=1
\usepackage[T1]{fontenc}
\usepackage[latin9]{inputenc}
\usepackage{bm}
\usepackage{amsmath}
\usepackage{graphicx}
\usepackage{amssymb}
\newcommand{\beq}{\begin{equation}}
\newcommand{\eeq}{\end{equation}}
\newcommand{\beqa}{\begin{eqnarray}}
\newcommand{\eeqa}{\end{eqnarray}}

\makeatletter

\@ifundefined{textcolor}{}
{%
 \definecolor{BLACK}{gray}{0}
 \definecolor{WHITE}{gray}{1}
 \definecolor{RED}{rgb}{1,0,0}
 \definecolor{GREEN}{rgb}{0,1,0}
 \definecolor{BLUE}{rgb}{0,0,1}
 \definecolor{CYAN}{cmyk}{1,0,0,0}
 \definecolor{MAGENTA}{cmyk}{0,1,0,0}
 \definecolor{YELLOW}{cmyk}{0,0,1,0}
 }

\makeatother

\begin{document}
\bibliographystyle{naturemag}

%\title{Absence of localization in T broken partite lattices}
\title{Anomalous Hall metals from strong disorder in class A systems on partite lattices}

\author{ Eduardo V. Castro}
\email{eduardo.castro@tecnico.ulisboa.pt}
\affiliation{CeFEMA, Instituto Superior T\'{e}cnico, Universidade de 
Lisboa, Av. Rovisco Pais, 1049-001 Lisboa, Portugal}
\affiliation{Beijing Computational Science Research Center, Beijing 100084, China }

\author{ Raphael de Gail}
\affiliation{Instituto de Ciencia de Materiales de Madrid, CSIC,
Sor Juana In\'es de la Cruz 3,  Cantoblanco,
E-28049 Madrid, Spain}

\author{M. Pilar L\'opez-Sancho}
\affiliation{Instituto de Ciencia de Materiales de Madrid, CSIC,
Sor Juana In\'es de la Cruz 3,  Cantoblanco,
E-28049 Madrid, Spain}

\author{Mar\'{\i}a A. H. Vozmediano}
\affiliation{Instituto de Ciencia de Materiales de Madrid, CSIC,
Sor Juana In\'es de la Cruz 3,  Cantoblanco,
E-28049 Madrid, Spain}

\date{\today}

\maketitle

{\bf Topological matter is a trending topic in condensed matter: From a fundamental point of view it has introduced new phenomena and tools, and for technological applications, it holds the promise of basic stable quantum computing. Similarly, the physics of localization by disorder, an old paradigm  of obvious technological importance in the field, continues revealing surprises when new properties of matter appear. This work deals with the localization behavior of electronic systems based on partite lattices with special attention to the role of topology. We find an unexpected result from the point of view of localization properties: A robust topological metallic state characterized by a non--quantized Hall conductivity arises from strong disorder  in class A (time reversal symmetry broken) insulators. The key issue is the nature of the disorder realization: selective disorder in only one sublattice in systems based on bipartite lattices. The generality of the result is based on the partite nature of most recent 2D materials as graphene or transition metal dichalcogenides, and the possibility of the physical realization of the particular disorder demonstrated in \cite{UBetal10}. An anomalous Hall metal arises also when the original  clean insulator is topologically trivial.} 

\section{Introduction}
\label{sec_intro}

After the seminal work of P. Anderson \cite{Anderson58} it was understood that in a non-interacting two dimensional electron system at zero temperature in spacial dimension $D\leq 2$ and in the thermodynamic limit, the electronic wave function will be localized by disorder. In more realistic situations the scaling theory of localization allowed a classification of the localization behavior of materials into universality classes set by symmetry and space dimensionality  \cite{AALR79,ATAF80} based on the Altland-Zirnbauer sets of random matrices  \cite{AZ97}. The advent of topological insulators \cite{BHZ06,HK10,QZ11} provided a new class of delocalized states, the edge states, robust under disorder provided some discrete symmetries were preserved. The symmetry classes were then adapted to include the topological features and a ``tenfold way" classification  was set \cite{SRFL08,EM08}. 

Centering the attention on the three non--chiral symmetry classes of the original Wigner--Dyson classification in two dimensions, we expect the following situation: 
All states will be localized in the orthogonal class AI (time reversal symmetry ${\cal T}$ with ${\cal T}^2=1$ preserved);  a mobility edge \cite{LR85}, i.e., a well defined energy separating a region of extended states from the localized states, is expected in the symplectic class AII (time reversal symmetry with ${\cal T}^2=-1$ preserved). Finally, in the unitary class A (${\cal T}$ broken), extended states can remain at particular energies. Only classes A and AII support topological indices. The prototypical examples in classe A are systems showing the integer quantum Hall effect (IQHE) and anomalous quantum Hall systems, the later exemplified by the Haldane model \cite{H88}.  Spin Hall systems \cite{KM05,KM05b} belong to class AII.

The interplay of topology and localization was first analyzed in the context of the robustness under disorder of the Hall conductivity quantization in the IQHE\cite{Pruisken88,Pru88,CC88,Letal94}. This is an example of a Chern insulator that belongs to  symmetry class A (all discrete symmetries are broken) in the standard classification. The mechanism for localization in both topological classes A, and AII, is  referred to as ``levitation and annihilation" \cite{OAN07}. For moderate disorder, the states in the edges of the conduction and valence bands  start to localize. As disorder increases, the gap is totally populated by localized states and the extended states carrying the Chern number shift towards one another and annihilate leading to the topological phase transition. The difference between the two classes is that, while in the symplectic class AII, a finite region of extended states with a well defined mobility edge  remains until the transition takes place,  there is no mobility edge in the unitary class A systems. The extended states carrying the Chern number are located at particular single energies.

We use the Haldane model \cite{H88} as a typical example of class A system based on a bipartite lattice. As it is known, depending on the parameter values, the model can represent a Chern or a trivial insulator.  The main result of this work is the finding of an extended region of delocalized states with a well defined mobility edge for strong disorder in class A systems when disorder is selectively distributed in only one sublattice. Why this is surprising is because there is no mobility edge in this class. Hence our result implies that the standard classification has to be modified. Moreover the final metallic state is an anomalous Hall metal even in the case when the clean starting system is a topologically trivial insulator.

In addition to its fundamental interest, the physics of this work can be realized in actual material systems. Many of the 2D materials relevant for technological or fundamental physics are based on bipartite lattices. Most prominent examples like  graphene  and its siblings silicene, germanene, or stanene;   black phosphorus, boron nitride, or  transition metal dichalcogenides MX$_2$ (M$=$ Mo, W and X$=$ S, Se) are based on the Honeycomb lattice \cite{Gibney15}.  The experimental possibility of inducing disorder selectively in one sublattice only, has been proven in \cite{UBetal10}.

\section{Model and methods}
\label{sec_model}
\begin{figure}
\begin{centering}
\includegraphics[width=0.7\columnwidth]{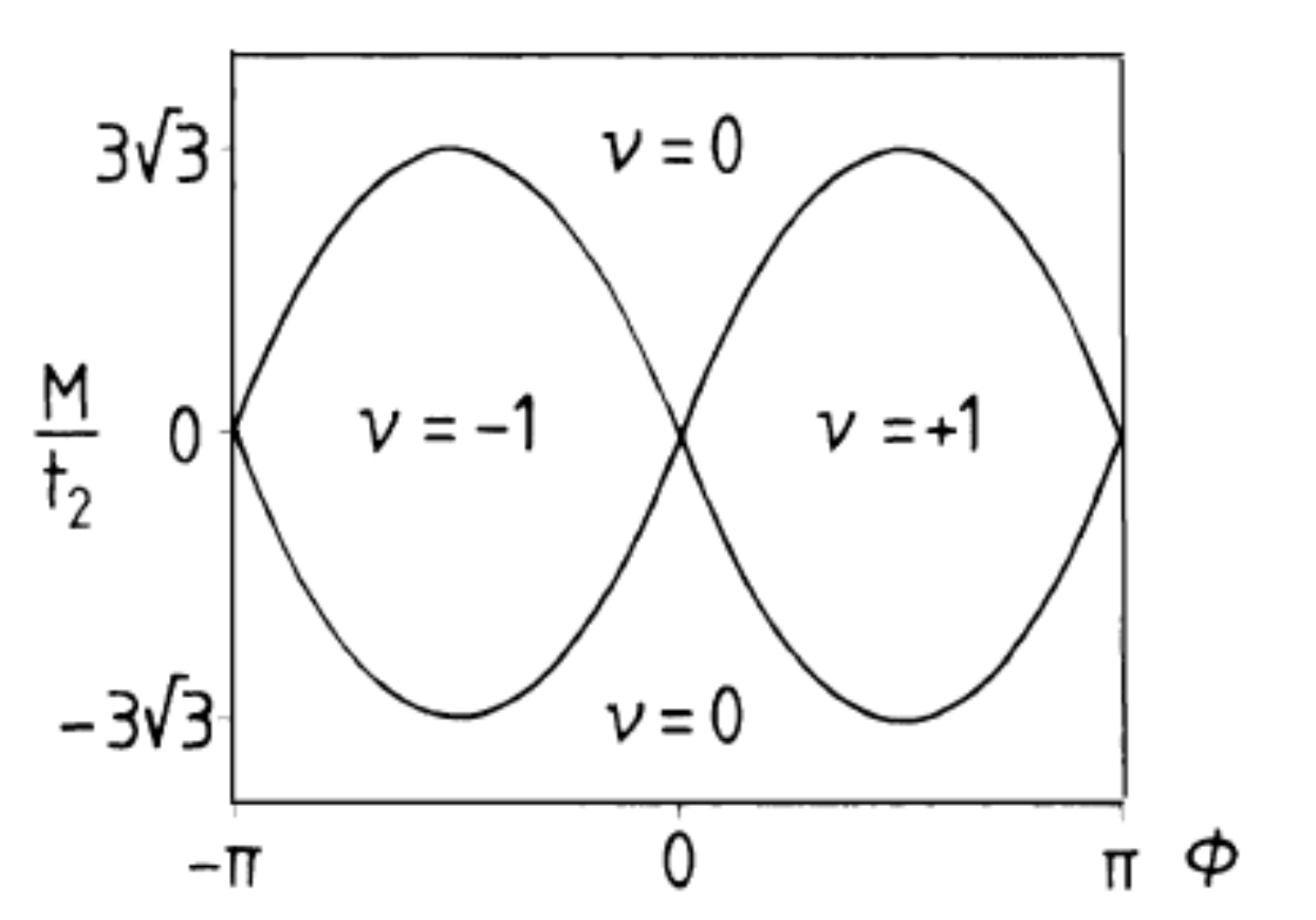}
\par
\end{centering}
\caption{\label{Fig_Hal}Phase diagram of the Haldane model as a function of the parameters
M and $\phi$ with $\vert t_2/t_1\vert <1/3$. The condition $\vert M\vert =3\sqrt{3} t_2\sin\phi$ sets the boundary between a trivial  (Chern number $\nu$=0 ) and a topological insulator $\nu=\pm 1$. }
\end{figure}
We use the Haldane model \cite{H88} as a generic example of a Chern topological insulator. 
The Haldane model tight binding Hamiltonian can be written as 
%cast in the
%form (see Figure~\ref{fig:lattice}),
\beqa
H  &=  -t\displaystyle\sum_{<ij>}c_i^{\dagger}c_j+%\\ \nonumber
-t_2\sum_{<<ij>>}e^{-i\phi_{ij}}c_i^{\dagger}c_j\\ \nonumber
&+  M\displaystyle\sum_i \eta_i c_i^{\dagger}c_i +\mbox{H.c.},
\label{TBHmodel}
\eeqa
where $c_i=A,B$ are defined in the two triangular sublattices that form the honeycomb lattice. 
The first term $t$ represents a standard
real nearest neighbor hopping that links the two triangular sublattices.
The next term represents a complex next nearest neighbor hopping $t_2 e^{-i\phi_{ij}}$
acting within each triangular sublattice with a 
phase  $\phi_{ij}$ that has
opposite signs $\phi_{ij}=\pm\phi$ in the two sublattices.
This term breaks time--reversal symmetry and opens a non--trivial topological gap
at the Dirac points. 
The last term represents a staggered potential($\eta_i=\pm 1$). It breaks 
inversion symmetry and opens a trivial gap at the Dirac points. The topological transition occurs at $\vert M\vert =3\sqrt{3} t'\sin\phi$ as indicated in Fig. \ref{Fig_Hal}.
We have done our calculations for the simplest case $\phi=\pi/2$ and a typical value $t_2$=0.1t. M has been set to zero except when analyzing the topologically trivial case.
A physical realization of the model with optical lattices has been presented in 
\cite{JMetal14}(see also \cite{W13}).
 
Potential (Anderson) disorder is implemented by adding to the Hamiltonian the term $\sum_{i\in A,B}\varepsilon_{i}c_{i}^{\dagger}c_{i}$,
with a uniform distribution of random local energies, $\varepsilon_{i}\in[-W/2,W/2]$. We will discuss two cases: disorder equally or selectively
distributed among the two sublattices.  For selective disorder the sum runs only over one sublattice. 

The Haldane model belongs to symmetry class~A  where the different topological
phases can be characterized by a $\mathbb{Z}$--topological number,
the Chern number $\nu$.
In the clean insulating system it can be computed from the single particle Bloch states $u_n({\bf k})$ as:
\beq
\nu_n=\frac{1}{2\pi}\int_S\Omega_z^n(k) dS,
\eeq
where the integral is over the unit cell and  $\Omega_z^n(k)$ is the $z$
component of the Berry curvature: 
$ \Omega^n(\mathbf{k})= \nabla_k\wedge A_n(\mathbf{k})$  defined from the
Berry connection:
${\cal A}_n({\bf k})=\left<u_n({\bf k})\vert -i\nabla_{{\bf k}}\vert u_n({\bf k})\right>$.
The non trivial topology of metallic states (anomalous Hall systems) is associated to a finite, non quantized Hall conductivity that can be computed using a Kubo formula.
The main technical difficulty in addressing disordered systems is the breakdown of translational symmetry which prevents working directly in momentum space. The subject being very old, many numerical and analytical tools have been worked out to deal with this oddity. Topological systems share the same problem as most topological indices are naturally defined in k space. We have used a numerical recipe based on the Kubo formula a to compute the Hall conductivity in the disordered tight binding model similar to that described in \cite{CB01}.

The localization behavior of the system has been explored with standard tools: Level spacing statistics, and inverse participation ratio (IPR) \cite{EM08}. A  transfer matrix method \cite{Abrahams10} has been also used to compute the localization length and confirm the presence of a mobility edge.  
% A list them with a brief description and the main references are presented in the supporting information.

\section{Warming up: Disorder equally distributed in both sublattices}
\label{sec_even} 

\begin{figure}
\begin{centering}
\includegraphics[width=0.7\columnwidth]{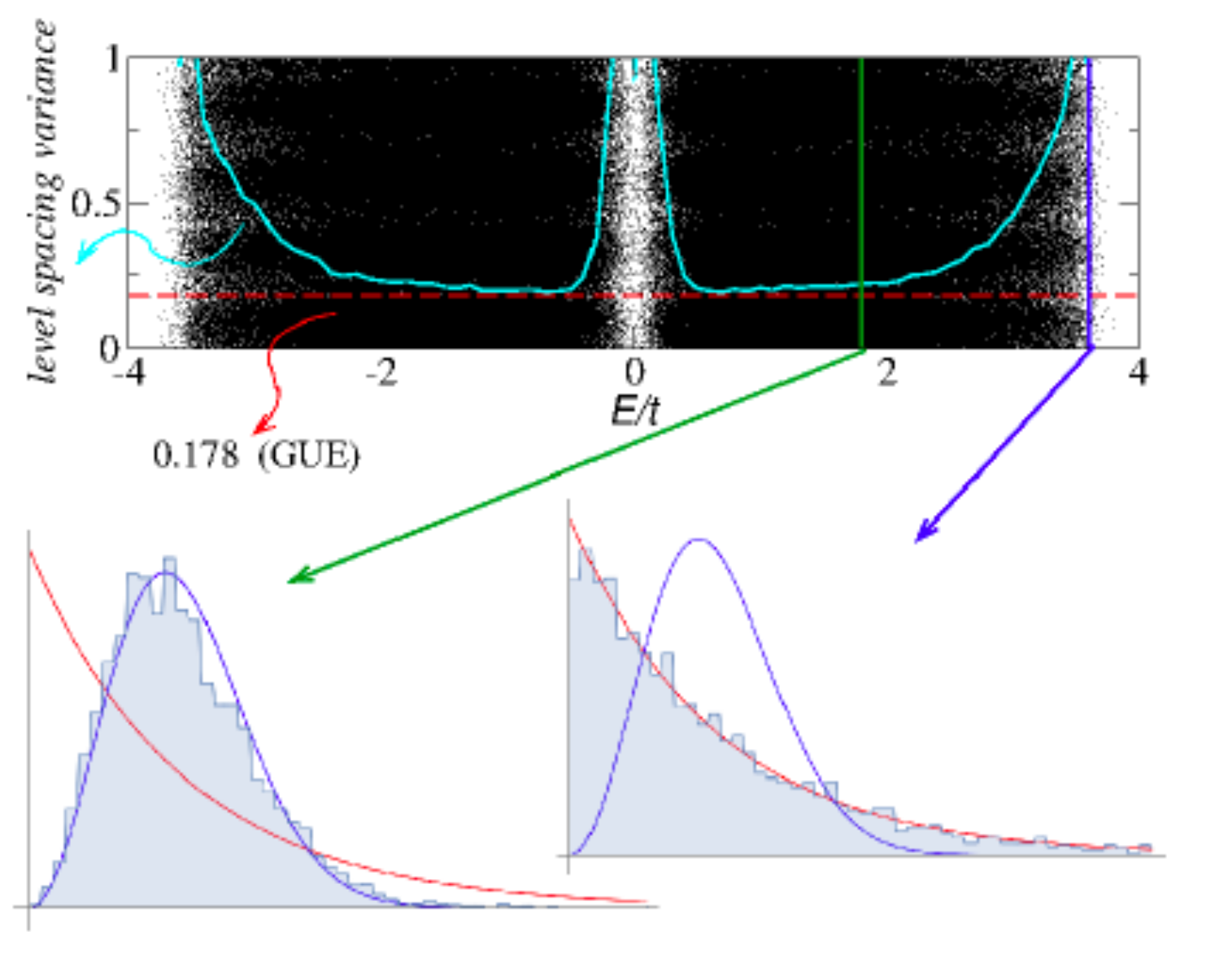}
\par\end{centering}
\caption{\label{Fig_equal} Level statistics analysis for the Haldane model with
Anderson disorder W= 3t equally distributed over the two sublattices. States are localized (Poisson distribution) all along the energy range. Extended states (GUE) are found at the two singular energies marked in the figure (green lines). This result agrees with the analysis done in Ref. \cite{Prodan11}. }
\end{figure}

\begin{figure}
\begin{centering}
\includegraphics[width=0.9\columnwidth]{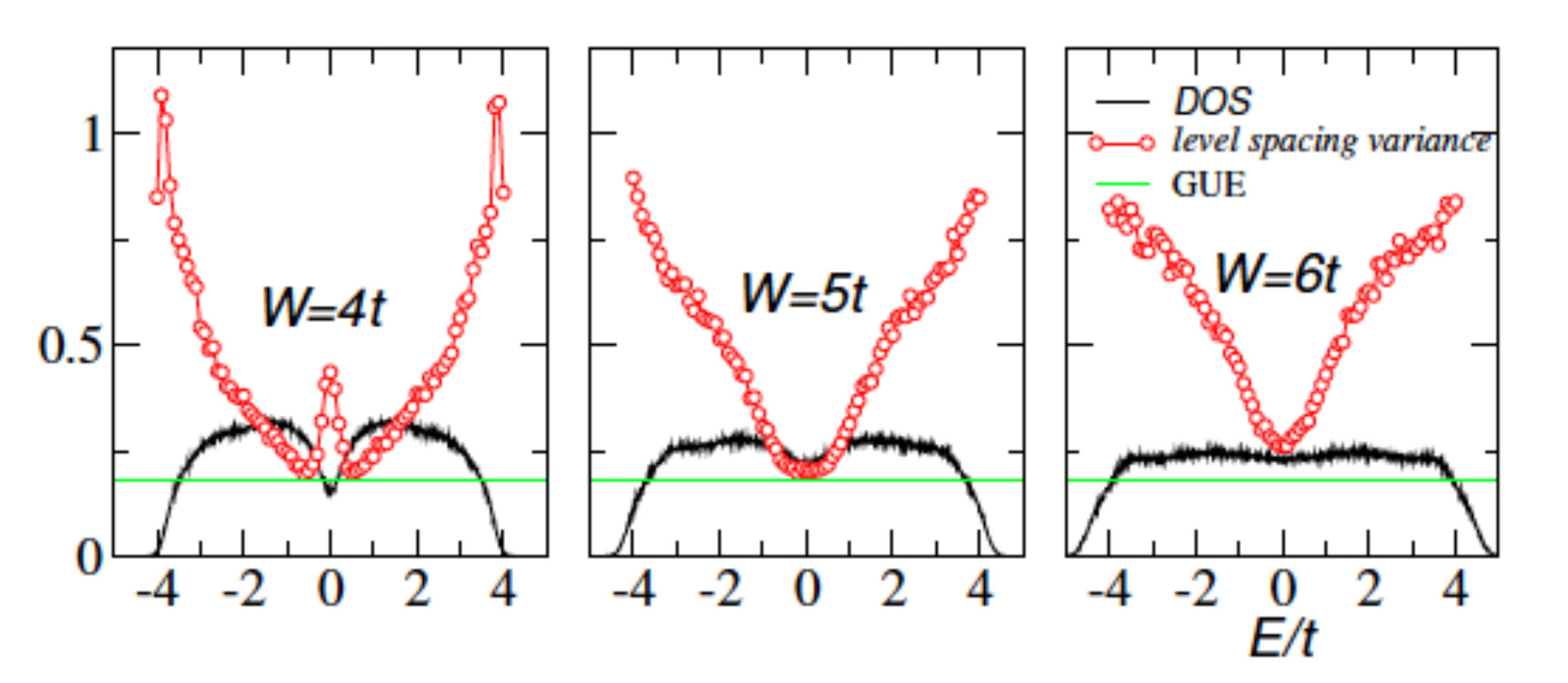}
\par\end{centering}
\caption{\label{Fig_equal2} Localization transition studied through level statistics
for the Haldane model with Anderson disorder equally
distributed over the two sublattices. The horizontal green line marks the GUE variance. The two extended states present at W=3t merge around W=5t and annihilate as disorder increases. All states become localized for $W>W_c\sim5t$. }
\end{figure}

We first present the case of Anderson disorder equally distributed in the two sublattices which shows the standard behavior of class A systems: extended states carrying the topological index remain at singular energies, approach each other as disorder increases (levitation) and merge (annihilation of the topological index). 
Figure \ref{Fig_equal} shows
the spectrum for the Haldane model with Anderson disorder
equally distributed over the two sublattices for a disorder strength W = 3t.
The dots correspond to a given eigenenergy for a given
disorder realization in a finite lattice with size d = 30. We
have performed 1000 disorder realizations. Superimposed
to the spectrum we show the level spacing variance as a
function of energy. The variance of the level spacing variation
contains information on the localization of the states at a given energy region
(details can be found in the supporting information).
Gaussian Unitary Ensemble (GUE) and Poisson (P) ensemble statistics are associated to extended  and localized states 
respectively. %Only for extended states we expect the variance to follow the prediction of the Gaussian Unitary Ensemble (GUE) for the statistics of the level spacing $s = \epsilon_{i+1}-\epsilon_i$. The probability distribution for GUE is $p(s) =\frac{32}{\pi^2}s^2\exp{-4x^2/\pi}$), and the respective variance $\left\langle s^{2}\right\rangle -\left\langle s\right\rangle ^{2}=0.178$. 
It is clear that there are two extended
states, one below the gap and another one above,
where the variance clearly approaches the GUE variance 0.178. 
%Note that the probability distribution p(s) is well approximated by a the Poisson distribution $p(s)=\exp{-s}$ away from the extended states (we exemplify with E = 3.6t, blue line), while it approaches the GUE for energies closer to the delocalized states (example in the figure for E = 1.8t, green line). 
These results
are in perfect agreement with those presented in Ref. \cite{Prodan11}.
In Fig. \ref{Fig_equal2} we show the level statistics variance and the
DOS for three different disorder strengths: $W = 4t, 5t, 6t$.
Levitation and annihilation is clearly operative, and the
critical disorder for localization is in good agreement with that obtained in ref. \cite{CLV15}
for the topological transition, $4t < W_c < 5t$.

\section{Main results}
\label{sec_results}

\subsection{Disorder selectively distributed in only one sublattice: Topological model}
\label{sec_selectivetopo}

\begin{figure}
\begin{centering}
\includegraphics[width=0.7\columnwidth]{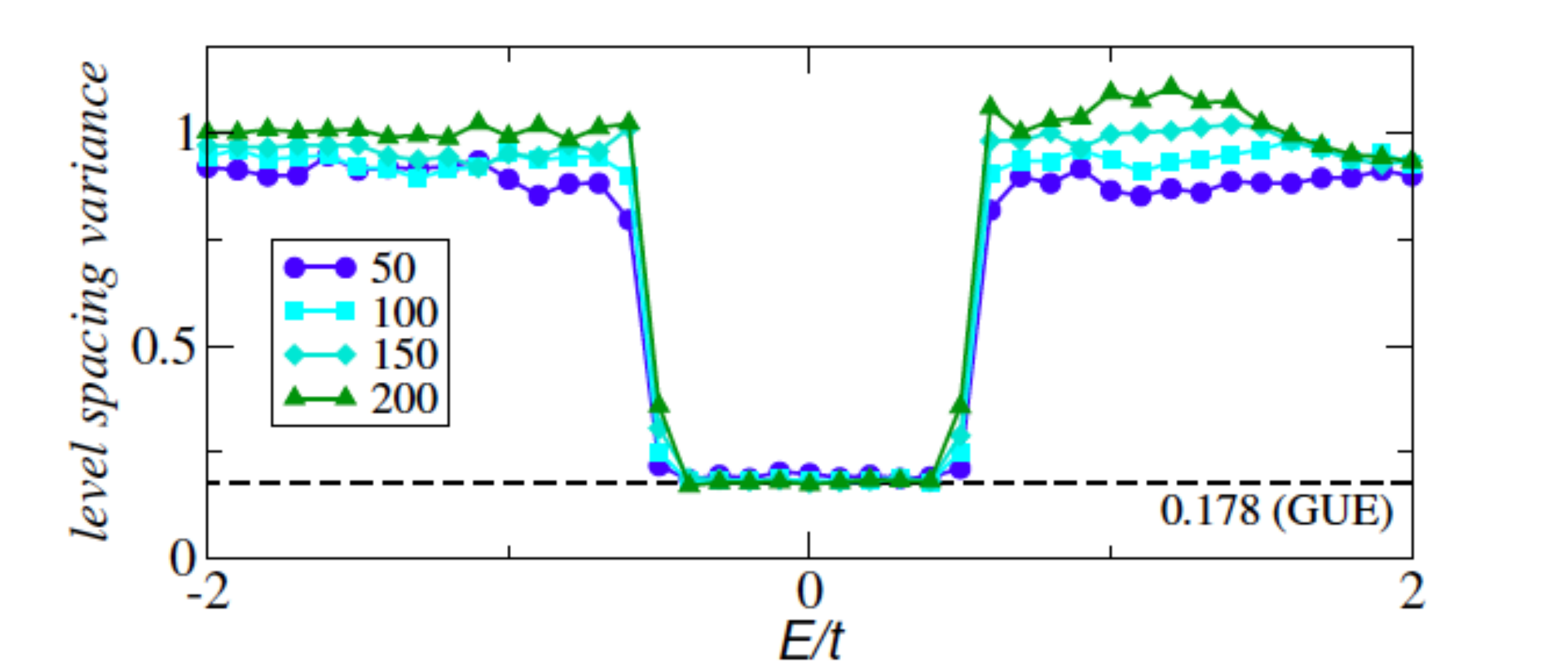}
\par\end{centering}
\caption{\label{selective2} Level spacing variation for increasing disorder 
strengths W/t in the selectively distributed disorder case.  
The middle region has the same variance as that of
GUE and corresponds to extended states. Even
though the transition is becoming sharper, the
region is not shrinking. This is a clear evidence for the existence of an
extended region of delocalized states. A mobility edge in the center of the band has emerged 
from the singular, isolated energies by increasing disorder.}
\end{figure}

The first unexpected result obtained is that, for selectively distributed disorder in only one sublattice, the class A system ends up in a robust metallic state where the extended states are separated from the localized states at the band edges by a well defined mobility edge.  %We use Anderson random potential parametrized through a local on-site energy $\epsilon_i$ equally distributed in the range $\epsilon_i \in [-W/2,W/2]$. 
\begin{figure}
\begin{centering}
\includegraphics[width=0.7\columnwidth]{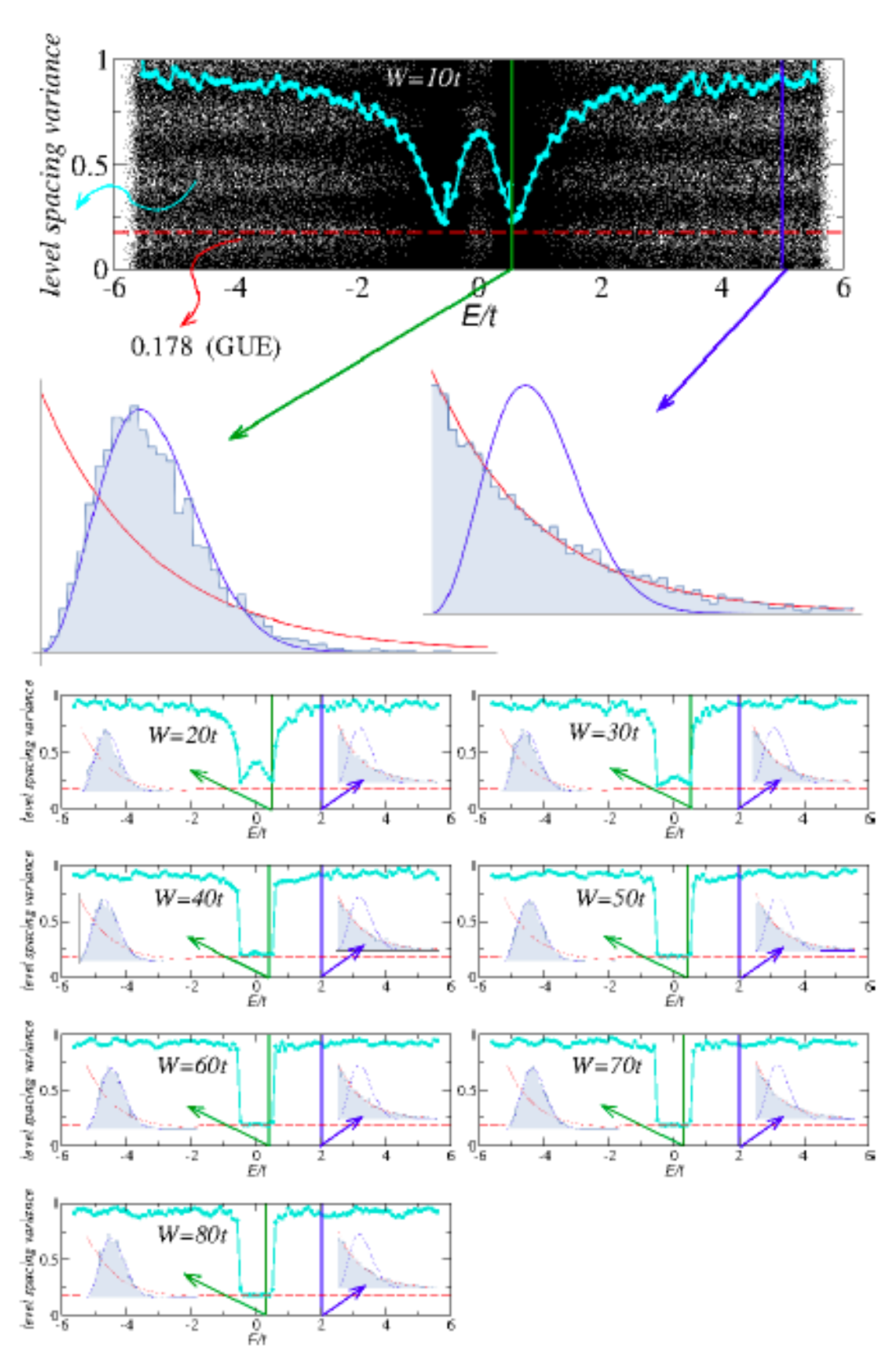}
\par\end{centering}
\caption{\label{Fig_selective} Level statistics analysis for the Haldane model with
Anderson disorder selectively distributed over one sublattice. The red dotted horizontal line marks the variance of the GUE associated to the presence of extended states. 
As disorder increases, the singular energies where extended states were located
at moderate disorder strength $W/t=20-30$, evolve to a full extended region of delocalized
states with a well defined mobility edge.}
\end{figure}
Figure \ref{Fig_selective} shows a level spacing statistic analysis of the
system for increasing disorder strength. What we see in the figure is the statistics associated to two characteristic energies in the  spectrum: one at the edge (blue line)
where states start to localize first when disorder is introduced, and one at the middle
of the band (green line) where extended states are expected to persist up to higher
disorder strength. The red dotted horizontal line marks the variance of the
GUE associated to the presence of extended states. 
We see that, as disorder increases, the singular energies where extended states were located
at moderate disorder strength $W/t=20-30$, evolve to a full extended region of delocalized
states with a well defined mobility edge.  Fig. \ref{selective2}
shows that the extended region of delocalized states is a robust feature that persists up
to a disorder strength of W=200t.
We have also set up a calculation of the localization length via a transfer matrix method  to confirm the presence of the mobility edge.

\begin{figure}
\begin{centering}
\includegraphics[width=1.0\columnwidth]{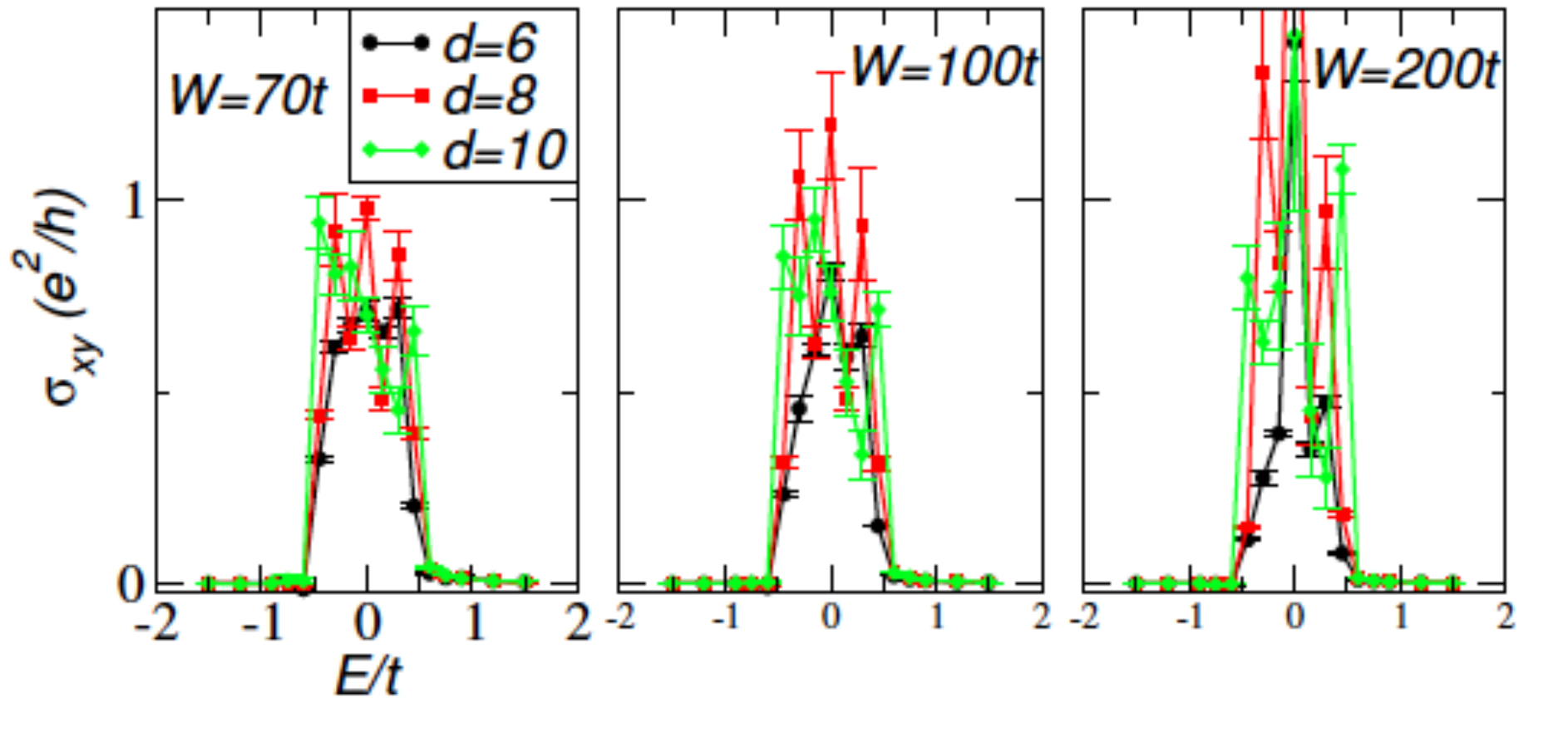}
\par\end{centering}
\caption{\label{HallC}Hall conductivity of the resulting metallic state emerging from the Chern insulator for disorder strengths above the critical value for the topological transition. 
 The conductivity is not quantized and depends on the chemical potential. Despite the big numerical error bars a finite non-zero value can be granted.}
\end{figure}
The topological nature of the metallic state is reflected in the calculation of the Hall conductivity shown in Fig. \ref{HallC}. 
In our previous publication \cite{CLV15} we showed that 
the Chern insulator suffered a topological transition to a trivial state at a critical disorder strength around $W_c\sim 50t$. What we see here is the further evolution to an anomalous Hall metal when disorder is further increased and the metallic state is well established. 
The panels in Fig. \ref{HallC} show that the Hall conductivity stays finite in
the metallic region for $W > W_c$. The different curves correspond to different sizes of the system. We see that for 
increasing system sizes $\sigma_{xy}$ is not decreasing what proofs that we are not dealing with a  finite size effect. Despite the large numerical error bars (bigger for smaller sizes), a finite conductivity can be granted.

\subsection{Disorder selectively distributed in only one sublattice: Topologically trivial case}
\label{sec_selectivetopo}

\begin{figure}
\begin{centering}
\includegraphics[width=1.0\columnwidth]{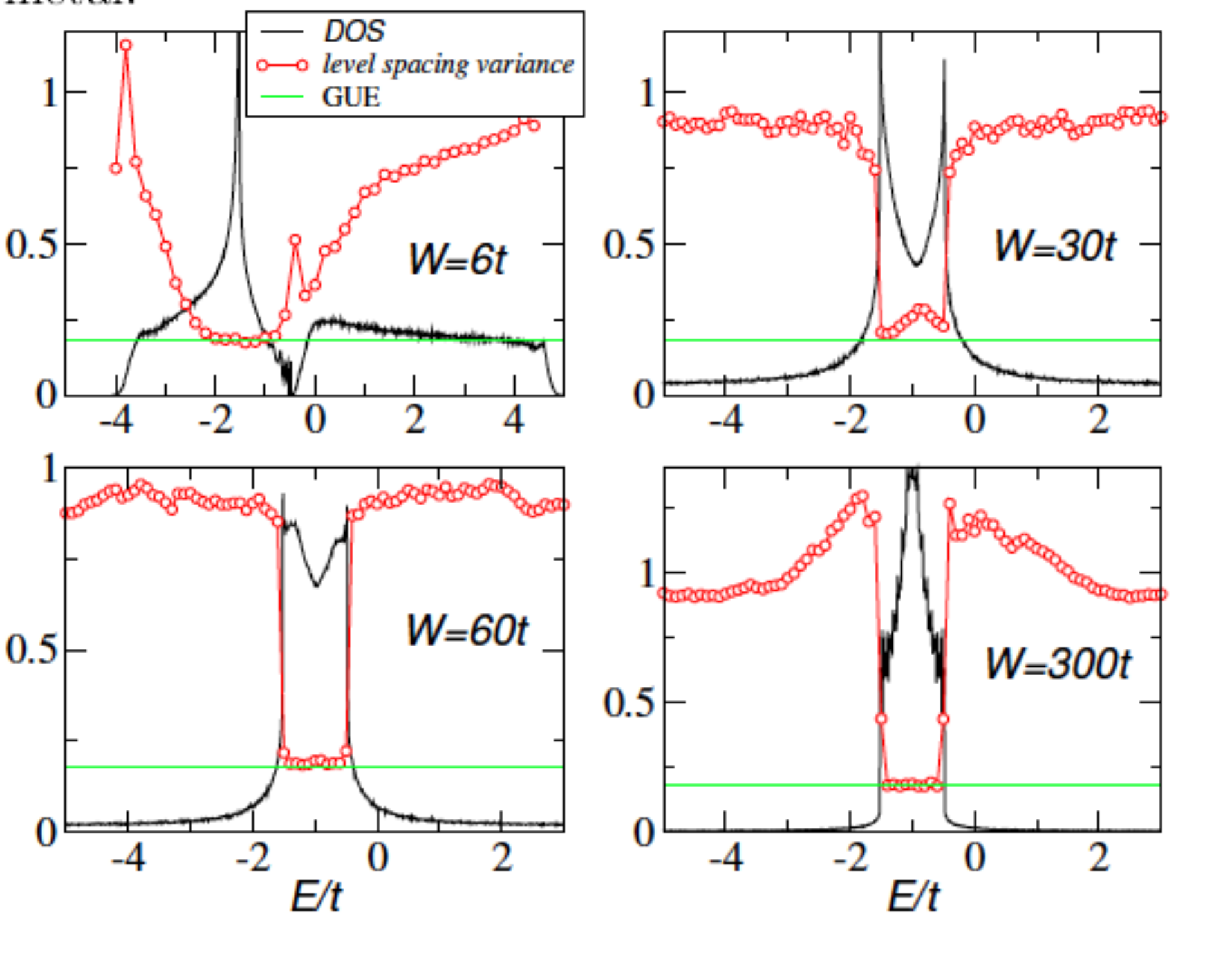}
\par\end{centering}
\caption{\label{Fig:TrivialVariance}Level spacing variation for the disordered trivial insulator in the selectively distributed disorder case. The results are very similar to these in the topological case in Fig. \ref{selective2}. }
\end{figure}
\begin{figure}
\begin{centering}
\includegraphics[width=1.0\columnwidth]{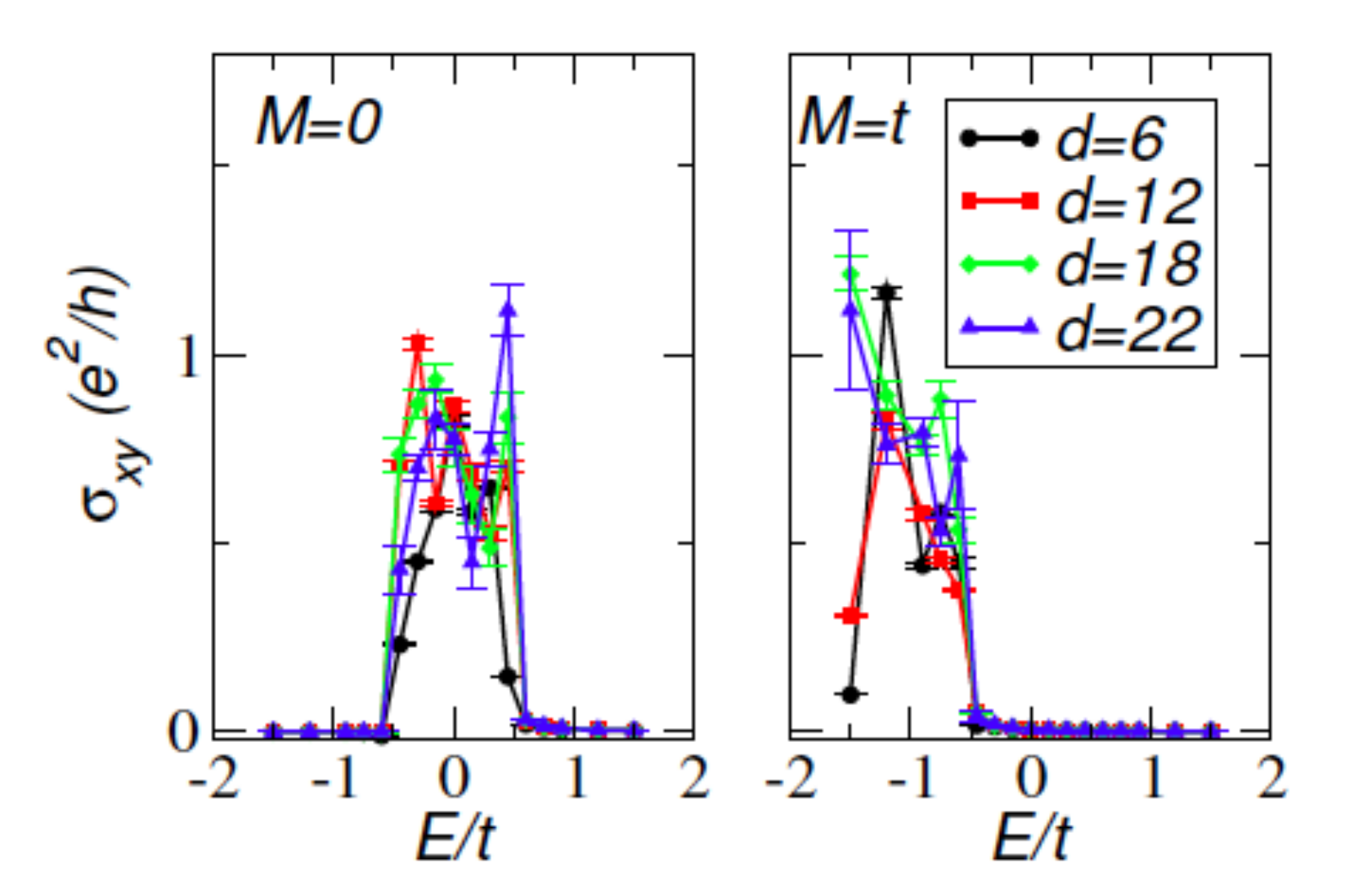}
\par\end{centering}
\caption{\label{Hall_trivial}A comparison of the Hall conductivity of the resulting metallic state emerging from the topological (left hand side) and trivial insulator (right and side). A finite Hall conductivity is obtained in both cases.}
\end{figure}
The second unexpected result is obtained when analyzing the trivial case. As it was mentioned when
describing the Haldane model, the parameters can be tuned to describe a trivial insulator for values of the staggered potential $\vert M\vert > 3\sqrt{3} t_2\sin\phi$. In order to ascertain the possible role of topology in the development of the metallic phase, we have analyzed the localization behavior of the trivial case with  generic values of the parameters chosen so that {\cal T} is still broken but the original insulator has a trivial gap with 
zero Chern number. The localization properties are shown in fig. \ref{Fig:TrivialVariance}. The final state is metallic with a well defined mobility edge. The topological nature of the final state is reflected in  the Hall conductivity  shown in  Fig. \ref{Hall_trivial}. 
We compare the topological $M = 0$ case  with the trivial $\vert M\vert > 3\sqrt{3} t_2\sin\phi$ for $t_2/t=0.1 e^{i\pi/2}$.
We have used larger system sizes,
and higher number of disorder realizations; 105, 105, 104,
and 5000 disorder realizations, respectively for d = 6,
12, 18, and 22. Note that $\sigma_{xy}$ 
is finite only in the
energy region where states have a larger amplitude in
the non-disorder sublattice. These findings agree well
with the results for the level spacing variance discussed
above.

%In order to understand the nature of the  the metallic phase obtained in this case, we have performed a series of calculations. The main results are shown in Figure \ref{Fig_selective300}. The left hand side depicts the DOS of the Haldane model in the topological phase with the standard values of the parameters used through the work: t'=0.1t, M=0, $\phi=\pi/2$ for increasing values of Anderson disorder distributed only in the B sublattice. Comparing with the case of equally distributed disorder (Figure \ref{Fig_equal}) we see that the two peaks structure characteristic of the Honeycomb lattice survive to a much higher values of disorder strength. For the limiting case W=300 t the DOS resembles that of the triangular lattice (see appendix \ref{sec_triangular}). First we have analyzed the partial IPR (see Appendix~\ref{sec_ipr}) on the disordered sublattice. As mentioned before, the particular value $\phi=\pi/2$ for the phase of the NNN hoppings in the Haldane hamiltonian makes the model particle--hole symmetric and, hence, in symmetry class D. For this class it is known that a metallic phase shows up with increasing disorder \cite{MFM15}. Anderson disorder breaks this symmetry  but on average it is restored. To make sure that we are discussing class A physics and not D, we have performed some calculations  adding  a finite real part to the NNN hopping in the Haldane model and checked that the conclusions are robust. These results are presented in Appendix \ref{sec_AD}.

%

\section{Understanding what is going on: Side questions.}
\label{sec_understanding}
This work rises a number of additional questions. We have addressed some of them, others remain open. 

For simplicity, a purely imaginary value of  the $t_2$ parameter $t_2=i 0.1 t$ has been used through the work ($\phi=\pi/2$). This choice induces an accidental particle--hole symmetry to the system that technically belongs to class D. We have performed some additional calculations with a non--zero real part for $t_2$ to ensure that we are discussing class A physics and found no qualitative differences. It could be thought that, since disorder affects only one sublattice, the final metallic state coincides with the trivial metal of the triangular lattice. The result of the Hall conductivity makes obvious that this is not so. We have performed some calculations of partial IPR and saw that the wave function of extended states has always some weight in the disordered sublattice. It is interesting to note that this is so also in the case of vacancy disorder when the disordered sublattice is depleted.  

%The fact that the final metal is an anomalous Hall metal even when the starting point is the trivial insulator shows the difference between the quantized and anomalous metallic Hall states. The main characteristic of class A systems is the breakdown of time reversal symmetry. While this is not enough to guarantee a topological insulator in the gapped case, it ensures an anomalous Hall conductivity when the system becomes metallic. In the trivial insulator phase of the Haldane model the (quantized) Chern number is zero as a result of a cancellation between bands with non homogeneous Berry curvatures.

Irrespective of the topological character of the clean system, the final state in the class A analyzed is a {\cal T} broken disordered metal with finite Hall conductivity. Our results show that, while a topological insulator can be become trivial by an appropriate tuning of the parameters as happens in the Haldane model,  the anomalous Hall metal is a very robust and a stable phase for  {\cal T} broken metals in the absence of an external magnetic field.

As we mentioned above, even when for the highest values of disorder, the disordered sublattice is never decoupled from the ordered one. The physics that we observe all along the work is that of the disordered Honeycomb lattice, as proven by the fact that the extended states found in the extreme disorder case have a non zero weight in the disordered sublattice. Even though the weight in the disordered sublattice is orders of magnitude smaller than in the clean sublattice, the two sublattices do not ``decouple". This explains the metallic nature of the final state: An electron in the disordered sublattice can propagate to a very long distant site by hopping to the clean sublattice, propagate there, and hop back. The probability of the process is suppressed to be of order $\alpha t^2$  but it is never zero. This also explains why the final metallic state is a topological metal since the anomalous Hall effect is due to the
interband matrix elements of the current operators \cite{KL54}. 
%The sensitivity of the localization transition to the sublattice selection  has been addressed in \cite{LHM14,OPK14} but no results similar to those described here were reported. A recent calculation of the conductivity tensor in the disordered Haldane model suggests that an asymmetry between sublattices A and B can help to stabilize the Chern insulator \cite{GCR15}, an intriguing result compatible with our findings.
   
The different behavior of  IQHE and anomalous Hall metals under disorder has been examined  in refs. \cite{ON02,ON03}.  The properties of  disordered topological metals arising from clean topological insulators have been analyzed in ref. \onlinecite{MR13}.
The metallicity of the final state  seems to be at odds with the non-linear
sigma model results \cite{EM08,MFM15} so it would be very interesting to implement the selective disorder case in this approach.

The physics described in this work can be realized in topological materials based on other more complicated partite lattices \cite{XLZ15}. The results presented in this work are conceptually important, although we recognize that to implement the selectively distributed disorder might be a hard task. To this respect, we note that experiments have been done in graphene where defects are located selectively in one sublattice to check the magnetic properties of the system \cite{UBetal10}. Artificial \cite{GMetal12} or optical lattices \cite{JMetal14} are other possibilities to realize this physics. 
 
\begin{acknowledgments}
We gratefully acknowledge useful conversations with Alberto Cortijo, Bel\'en Valenzuela, Fernando de Juan, Adolfo G. Grushin,  and J. A. Verg\'es. 
EC acknowledges the financial support of FCT-Portugal through grant No. EXPL/FIS-NAN/1728/2013. This research was supported in part by the Spanish MECD grants FIS2014-57432-P, the  European Union structural funds and the Comunidad de Madrid MAD2D-CM Program (S2013/MIT-3007),   the European Union Seventh Framework Programme under grant agreement no. 604391 Graphene Flagship FPA2012-32828.
\end{acknowledgments}

\bibliography{Chern2}
\end{document}